\documentclass[preprint,preprintnumbers]{revtex4}

\usepackage{graphicx}

\usepackage{amssymb}

\usepackage{mathrsfs}

\begin{document}

\title{The  $S$-matrix for surface boundary states: an application to photoemission  for Weyl semimetals} 
\author{D. Schmeltzer}

\affiliation{Physics Department, City College of the City University of New York,  
New York, New York 10031, USA}

 \begin{abstract}

We present a new theory of photoemission for Weyl semimetals. We derive this theory using a model with  a boundary surface    at $z=0$. Due to the boundary, the self adjoint  condition needs to be verified  in order to ensure physical solutions. The solutions are  given  by  two chiral zero   modes which propagate on the boundary.
 Due to the  Coulomb interaction,  the chiral  boundary model is in the  same universality class  as interacting  graphene. The interactions cause  a temperature dependence of the  velocity and  and   life time.
\noindent
Using the principle of  minimal coupling, we identify the electron-photon  Hamiltonian.  The photoemission intensity  is  computed  using  the $S$-matrix formalism.  The   $S$-matrix is derived  using the initial photon state, the  final state of a  photoelectron  and a hole in the valence band.  The photoemission  reveals   the final  valence band  dispersion $ \hbar v(\pm k_{y}-k_{0})+\hbar\Omega$   after absorbing a  photon of frequency $ \Omega$ ($k_{0}$ represents the shift in the momentum due to the crystal potential).   The momentum in the $z$ direction is not conserved, and  is  integrated out. As a result, the scattering matrix is a function of the parallel  momentum . We observe two dimensional contours,  representing the     $^{''}$Fermi arcs $^{''}$, which for     opposite spin polarization  have opposite curvature. This theory is in   agreement with previous  experimental  observations.

 \end{abstract}

\maketitle

\textbf{I. INTRODUCTION}

\vspace{0.2 in}
Photoemision is the standard  method used to provide information on the band structure in metals \cite{Mahan}. No such theory exists for the Weyl semimetal  where the     low energy electrons couple to photons via $ \vec{\sigma}\cdot\vec{A}$ in the presence of a boundary  at  $z=0$. The  presence of the  boundary   demands special considerations  \cite{Witten}.

Weyl fermions represent a pair of particles with opposite chirality, described  by the massless solution of the Dirac equation \cite{Weyl}. Recently, it has been proposed that in materials with two non-degenerate bands crossing at the Fermi level in three dimensional $(3D)$ momentum space, the low energy  excitations can be described by the Weyl equations, allowing a condensed matter realization of Weyl fermion quasiparticles \cite{Vishwanath,Xu}. The band crossing points are called Weyl points,
and materials possessing such Weyl points are known as Weyl semimetals $(WSMs)$.
The bulk of the $WSMs$  is dominated by Weyl points with linear,  low energy  excitations. The Weyl points come in pairs with opposite chirality \cite{Ninomya}.
The surface state of the $WSMs$  is  characterized by $^{''}$Fermi arcs$^{''}$ that link the projection of the bulk Weyl points  in the  Brillouine zone.  In the presence of  parallel electric and magnetic fields, the $WSMs$ have a large negative  magnetoresistance \cite{Spiwak}, due to the Adler-Bell-Jackiw chiral anomaly \cite{Son}.
The $WSMs$  exist in materials where time-reversal symmetry or inversion are  broken \cite{Vishwanath}.
Recently, the non-centrosymmetric and non-crystal magnetic transition-metal monoarsenide/posphides  $TaAs$ ,$TaP$, $NbAs$ and $NbP$ have been predicted to be  $WSMs$ with $12$ pairs of Weyl points \cite{Science}
as  demonstrated  by  photoemission \cite{BQ}, scanning tunneling microscopy \cite{Stern} and measurements of the quasiparticle profile   \cite{Belopolski,Yazdani}. In contrast to the  large  amount of work devoted to photoemission experiments, few   theoretical results  exist   for Weyl materials \cite{Behrends}.

 The early photoemission  theories formulated by  \cite{Mahan} and applications to  experiments \cite{Eastman} were based on the  matrix element   $\vec{A}\cdot \vec{p}\approx \vec{A}\cdot i\frac{ \vec{\nabla}V(\vec{x})}{\Omega} $   between the wave functions    with the crystal potential  $ V(\vec{x})$ and  the free space (where $\Omega$   is the laser frequency). In  a regular metal, the dispersion of the electrons is approximately the same in  the crystal and the vacuum. The only difference is the crystal potential, which is zero outside the crystal.  As a result, the scattering problem involves the matrix element introduced by the crystal.  Such a calculation is probably not possible for a Weyl semimetal where  the  $\vec{A}\cdot \vec{p}$ is not valid. The  potential that  describes the transition from  free electrons to Dirac electrons with nodal points is not available. Morever,  the presence of  a boundary  further complicates the problem. For  a  Dirac operator,  the self-adjoining conjugation is  not automatically satisfied, and  according to  \cite{Witten}, the  boundary  condition  constrains the solutions.  

 We present a new theory which takes  into consideration the boundary condition and the Dirac dispersion. This will be done using a    non-linear  model for   $WSM$ with two Weyl points  restricted to  the region $z<0$ and a  free electron  system  in the region $z>0$ . In the region  $z<0$ the Hamiltonian is built from two terms $ h(\vec{k},z ) =h_{0}(\vec{k},z) +h^{\perp}(\vec{k})$.   $ h_{0}(\vec{k},z)$ contains the nonlinear part and  the surface  at  z=0. The second term, $ h^{\perp}(\vec{k})$,  is $z$- independent with the momentum     $\vec{k}$  parallel to the   surface $z=0$.  The parallel  momentum  of the emitted free electrons  is  equal to   the parallel momentum of    $WSM$  electrons,  the $k_{z}$ component is not conserved and is integrated out, leaving only the conserved momentum $\vec{k}$.  This Hamiltonian  is time-reversal invariant and has  broken inversion symmetry resulting in a  model   with  a monopole-anti-monopole pair.  The combination of the boundary condition and the Weyl nodes gives rise to a set of zero modes for the Hamiltonian $h_{0}(\vec{k},z)$. The maximal amplitude to find an electron on the    surface  at $ z=0$ is obtained for zero mode spinors .  The term  $h^{\perp}(\vec{k})$ gives rise  in the second quantized form to  the Hamiltonian   $H^{\perp}$ with  two chiral modes .

 For $z>0$, the model is given    by the free electron $ f_{\sigma}(\vec{k},z)$,  $ f^{\dagger}_{\sigma}(\vec{k},z)$ operator with the parabolic  energy dispersion  $E(\vec{k},k_{z})$.
 To ensure the   continuity  of the wave function  from  the bulk to the free space,    we will introduce the interface  Hamiltonian  $ H^{(interface)}$ .    The interface Hamiltonian describes the  region between the bulk electrons  and the  free space electrons  and   has  a  width $d$  around $z=0$.

\noindent 
In the presence of the photon  field  $\vec{A}$, the Hamiltonian    $H^{\perp}$ is replaced  by  $H^{\perp}_{0}$ +  $H^{\perp}_{A_{y}}$  where   $H^{\perp}_{A_{y}}$  is the photon Weyl-semimetal coupling. The photoemission  scattering Hamiltonian   $H^{ext.}_{int.}$ is built from the interface and the  photon-Weyl  coupling:
$ H^{ext.}_{int.}$ =  $H^{\perp}_{A_{y}}$+$H^{(interface)}$.

\noindent
The  Hamiltonian   $H^{ext.}_{int.}$ allows  us to compute the  $S$-matrix   \cite{Peskins}  where the incoming state $| i\rangle$  scatters to an   outgoing state  $| f\rangle$. The matrix $S_{f,i}$ is given by:

\noindent
$_{outgoing}\langle f| i\rangle  _{incoming}=\langle f |T\Big[e^{-\frac{i}{\hbar}\int_{-\infty}^{\infty}\,dt' H^{ext.}_{int.}(t')}\Big]|i \rangle=S_{f,i}$

\noindent
When   the initial state  $| i\rangle$ is an  incoming   photon,  the  transition to  the final state  $|f\rangle  $ is represented by the outgoing   emitted electrons  and a hole with a specific  chirality  is excited in the  valence band.   
 The formalism will be used  to determine   the valence band and conduction band  dispersion  and  to  demonstrate the presence of the $^{''}$Fermi  arcs$^{''}$.  Due to  the change in chirality when we cross the point $k_{x}=\pm M$, the arcs for  spin down electrons have opposite curvature to the ones with the spin  up electrons .

Relaxation effects are considered in  section $VI$ where Coulomb interactions are considered.

 In this paper we have obtained  the following  results:

(a) We have obtained  the chiral zero  modes for the Weyl boundary Hamiltonian.

(b)Using the minimal coupling principle, we have obtained the coupling between the photons and electrons.

(c) Using the $S$-matrix formalism \cite{Peskins},  we have obtained the intensity of the emitted photoelectrons from which the dispersion of the surface valence band  was  extracted.

(d) Based on  the   the scattering matrix  intensity,  we have    confirmed  that  the final  valence band  dispersion is $\epsilon=\hbar v(\pm  k_{y} -k_{0})+\hbar\Omega$,  after absorbing the photon of frequency $ \Omega$ .  $-\hbar v k_{0}$ is the potential with respect to the outside  crystal. 

(e)    The photon absorption  allows us to map  the conduction band dispersion.

(f) The momentum in the $z$ direction is not conserved and is integrated out.  This allows   us to project the intensity into two dimensional contours and observe the    $^{''}$Fermi arcs $^{''}$.

(g)  The change in   chirality when we cross the nodal  points  $k_{x}=\pm M$ determines the curvature of the  arcs. For  spin down electrons, the   curvature  of the arcs is opposite to the one with spin up electrons.

(h) Inclusion  of the Coulomb interaction shows that the chiral zero modes which propagate on the boundary are in the same universality class as interacting  graphene.  As a result, the velocity and the life time become temperature -dependent.

(k) This theory is in   agreement with the experimental  observation \cite{BQ}. 

The paper is structured as follows. In  Sec.II we present  the WSM model and discuss the chiral zero modes   resulting from   the boundary   conditions.
Sec.III is devoted to the identification of the solid vacuum  interface  Hamiltonian. In Sec.IV and in Appendix A we study the detection  of photoelectrons. In chapter V we present our  photoemission results and compare  them with the results given in the literature. Coulomb interactions are considered in section VI and Appendix B.
 Sec.VII is devoted to conclusions.

\vspace{0.2 in }

.

\textbf{II.  The Weyl  Hamiltonian  with a boundary  at  $z=0$  confined to the crystal  region $ -L\leq z \leq 0$ }

\vspace{0.2 in}

A WSM model  without a boundary and  two nodes $\vec{M} =[ \pm M ,0,0]$ is given by the Hamiltonian:
\begin{equation}
\tilde{H}=\int\,d^3x \hbar v\Big[\hat{\Psi}^{\dagger}_{R}(\vec{x})\vec{\sigma}\cdot\Big(-i\vec{\partial}-\vec{M}\Big)\hat{\Psi}_{R}(\vec{x})
-\hat{\Psi}^{\dagger}_{L}(\vec{x})\vec{\sigma}\cdot\Big(-i\vec{\partial}-(-\vec{M})\Big)\hat{\Psi}_{L}(\vec{x})\Big]
\label{equattion}
\end{equation}
We observe that  the  Hamiltonian  in Eq.(1) describes fermions with opposite chirality  and  two singularities at  $k_{x}=\pm M$.
This model is oversimplified and does not include the  band dispersion which     connects  the  two nodes. In order to observe this connection  we need  to study  a model  with two  non-linearly dispersed   bands. We are guided by the fact that  the singularities at  $k_{x}=\pm M$ describe a monopole and   anti-monopole . The monopole-anti-monopole  is  present   when one of the symmetries, time reversal or inversion symmetry, is broken.  To describe the crossing of the bands in momentum space, we will introduce a quadratic function of momentum   $g(k^2_{x}-M^2)$  which  reproduces  the nodes at $ \pm M$ ( this polynomial is obtained by replacing $-cos(k_{x}) +1 \approx \frac{k^2_{x} }{2}$) for   the two  band  Hamiltonian   $\hat{h}(\vec{k},z)$:
$\hat{h}(\vec{k},z)=\hbar v \Big[\sigma_{y}\tau_{3}k_{y}+\sigma_{z}\tau_{3}k_{z}+\sigma_{y}\tau_{2} g(k^2_{x}-M^2)\Big]$.

\noindent
This Hamiltonian  is invariant with respect to  the  time-reversal symmetry   and has a  broken inversion symmetry. As a result, 
 $k_{x}=\pm M$   is  a monopole-anti -monopole pair.

\noindent 
Next, we introduce a   Hamiltonian with a boundary surface  at $z=0$. This  is obtained by replacing the  momentum $ k_{z}$ with   $-i\partial_{z}$ and restricting the space to   $ -L\leq z \leq 0$ where the potential of the crystal with respect to the vacuum  is given by  $-V_{0}$. 
 We use the notation   $-V_{0}= -\hbar v k_{0}$ for  the $z\leq0$ region  and   $k_{0}=0$ for  $z>0$. The Hamiltonian with  the boundary at $z=0$ is given by:
\begin{eqnarray}
&&H=\int\frac{d^{2}k}{(2\pi)^2} \int_{-L}^{0}\,dz\Big[\hbar v\hat{\Psi}^{\dagger}(\vec{k},z)\Big(\sigma_{y}\tau_{3}k_{y}+\sigma_{z}\tau_{3}(-i\partial_{z})+\sigma_{y}\tau_{2} g(k^2_{x}-M^2)-k_{0}\Big)\hat{\Psi}(\vec{k},z)\Big]
\nonumber\\&&=\int\frac{d^{2}k}{(2\pi)^2} \int_{-L}^{0}\,dz\Big[\hbar v\hat{\Psi}^{\dagger}(\vec{k},z)h(\vec{k},z)\hat{\Psi}(\vec{k},z)\Big]\nonumber\\&&
\end{eqnarray}
The Hamiltonian in Eq.(2) consists of two orbitals  described by the   Pauli matrices $\tau_{1}$,$\tau_{2}$ and $\tau_{3}$. The spin  of the electrons is  introduced with the help of  the Pauli matrices $\sigma_{x}$,$\sigma_{y}$ and $\sigma_{z}$.

\noindent
 The translation symmetry along the $y$ direction is preserved, so that $k_{y}$ is a good quantum number.
For   the  semi-infinite  system   in  Eq.(2), we need to check  if the Dirac operator with the boundary term $\sigma_{z}\tau_{3}(-i\partial_{z})$  has  real eigenvalues.
We  apply the division described in   \cite{Shou} to the  the Hamiltonian $h(\vec{k},z)$:  
\begin{eqnarray}
&&h(\vec{k},z)=h_{0}(\vec{k},z)+h^{\perp}(\vec{k})\nonumber\\&&
h_{0}(\vec{k},z)=\hbar v \Big[\sigma_{z}\tau_{3}(-i\partial_{z})+\sigma_{y} \tau_{2}g(k^2_{x}-M^2)\Big]\nonumber\\&&
h^{\perp}(\vec{k})=\hbar v \Big[\sigma_{y}\tau_{3}k_{y}-k_{0}\Big]\nonumber\\&&
\end{eqnarray} 
 In  the region $-L\leq z\leq 0$, the Hamiltonian  $h_{0}(\vec{k},z) $    obeys   the eigenvalue equation:
\begin{eqnarray}
&&\Big[\sigma_{z}\tau_{3}(-i\partial_{z})+\sigma_{y}\tau_{2} g(k^2_{x}-M^2)\Big]U(\vec{k},z)=EU(\vec{k},z), \hspace {0.1 in}-L\leq z\leq0 \nonumber\\&&
\end{eqnarray}
 The eigenvector  $U(\vec{k},z)$ is a function of the      momentum   $\vec{k}$   parallel to the surface  $z=0$.
For the eigenvector  $U(\vec{k},z)$, we seek a   solution of form   $U(\vec{k},z)  =e^{\lambda z}V(\vec{k})$  and  find  the eigenvalues $ E=\pm\sqrt{\lambda^2- (g(k^2_{x}-M^2))^2}$. The self-adjoining condition  \cite{Witten} is satisfied  for real  eigenvalues $E$.  The  condition  for real eigenvalues $E$  is equivalent   with   $\lambda^{2}> (g(k^2_{x}-M^2))^2$. Since  the  electrons  are   confined to the region $-L\leq z\leq 0$, only solutions with positive  $\lambda$ are acceptable in  the  limit $L\rightarrow\infty $  (the   normalization of the   wave function  $U(\vec{k},z)$ demands that  $\lambda$  obeys the condition    $\lambda z<0$,  while  negative values of $\lambda$  are excluded in the limit $L\rightarrow\infty $). Using the relationship between $\lambda$ and the eigenvalue $E$, we find that the  amplitude of the wave function in the region    $ z\leq0$ is given by  $ e^{\lambda z} =e^{\sqrt{E^2+ (g(k^2_{x}-M^2))^2}z} $. Since $E$ must be $real$, we determine  that the   maximal amplitude   $ e^{\lambda z}$ is achieved    for  zero mode solutions  $E=0$, resulting in the  amplitude   $e^{|(g(k^2_{x}-M^2)|z}$.  As a result, the  maximal amplitude   of the wave function    $U(\vec{k},z) $  in the region    $ z\leq0$  is given by the zero mode solution:
\begin{equation}
 U(\vec{k},z)  =e^{\lambda z}V(\vec{k})=\theta[k^2_{x}-M^2]e^{g(k^2_{x}-M^2)z}\eta_{i,+}+\theta[-k^2_{x}+M^2]e^{-
g(-k^2_{x}+M^2)z}\eta_{i,-}(\vec{k}); \vspace{0.05 in} k^2\neq M^2 
\label{eqw}
\end{equation}
$\eta_{i,\pm}$, $ i=1,2$ are the two zero mode spinors, the index $\pm$ refers to the space region $k^2_{x}>M^2$ and $M^2>k^2_{x}$, respectively.
\begin{eqnarray}
&&\eta_{1,+}=\sqrt{\frac{1}{2}}\Big[i ,0,0,1\Big]^{T} ,\hspace{0.05 in} \eta_{1,-}=\sqrt{\frac{1}{2}}\Big[-i,0,0,1\Big]^{T} \nonumber\\&&
\eta_{2,+}=\sqrt{\frac{1}{2}}\Big[0,i,1,0\Big]^{T}, \hspace{0.05 in}   
\eta_{2,-}=\sqrt{\frac{1}{2}}\Big[0,-i,1,0\Big]^{T}\nonumber\\&&
\end{eqnarray}
This amplitude  will contribute the most  to the intensity of the emitted photoelectrons.

\noindent
 The fermion spinor $\hat{\Psi}(\vec{x},z)$ is replaced  by   the projected zero mode spinor   $\Psi(\vec{x},z)$. 
\begin{eqnarray}
&&\Psi(\vec{x},z)=\int\frac{d^{2}k}{(2\pi)^2} \sqrt{2\hat{g}s(k_{x})}e^{\hat{g}s(k_{x})z}e^{i\vec{k}\cdot \vec{x}}\sum_{i=1,2}\sum_{s=\pm}C_{i,s}(\vec{k})\alpha_{s}(k_{x}) \eta_{i,s} 
,\hspace{0.05 in} k^2_{x}\neq M^2\nonumber\\&&
\end{eqnarray}
The spinor  is normalized in the region $[-L,0]$. This gives rise to the factor $ \sqrt{2\hat{g}s(k_{x})}$, where the exponent  originate from the zero mode Eq.6.
The notation    $ \alpha_{s=1}(k_{x})\equiv\theta[k^2_{x}-M^2]$  is the step function which is $1$ for $k^2_{x}-M^2>0$ and zero for   $k^2_{x}-M^2<0$. Similarly,   $ \alpha_{s=-1}(k_{x})\equiv\theta[(-k^2_{x}+M^2]$  is $1$ for  $M^2 -k^2_{x}>0$ and zero for   $M^2- k^2_{x}< 0$.
 We simplify the notation  in Eq.(5),  $ g| (k^2_{x} -M^2) | \equiv \hat{g} s(k_{x})$ where $s(k_{x})$ and $\hat{g}$  are  given by, $s(k_{x})=|[(\frac{k_{x}}{M})^2-1]|$,  $\hat{g}=gM^2$, respectively.

Using the spinor representation given in Eq.(7),we diagonalyze the Hamiltonian $h^{\perp}(\vec{k})$. In the presence of the photon field $\vec{A}(\vec{x})$, we obtain:

\begin{eqnarray}
&&H^{\perp} =H^{\perp}_{0}+H^{\perp}_{A_{y}}=\int\frac{d^2k}{(2\pi)^2}\int_{-L}^{0}\,dz \hbar v \Psi^{\dagger}(\vec{k},z) h^{\perp}(k_{y}-A_{y}) \Psi(\vec{k},z)  \nonumber\\&&H^{\perp}_{0}=\int\frac{d^2k}{(2\pi)^2}\sum_{\pm}  \hbar v \Big[ k_{y}\Big(-iC^{\dagger}_{1,s}(\vec{k})C_{2,s}(\vec{k})\alpha_{s}(k_{x}) +iC^{\dagger}_{2,s}(\vec{k})C_{1,s}(\vec{k})\alpha_{s}(k_{x})\Big) \nonumber\\&&-k_{0}\Big(C^{\dagger}_{1,s}(\vec{k})C_{1,s}(\vec{k})\alpha_{s}(k_{x})+C^{\dagger}_{2,s}(\vec{k})C_{2,s}(\vec{k})\alpha_{s}(k_{x})\Big)\Big]   \nonumber\\&&
H^{\perp}_{A_{y}}=\sum_{s=\pm}\int\frac{d^2k}{(2\pi)^2}W(k_{x},\vec{q})A_{y}(\vec{q},t)\Big[iC^{\dagger}_{1,s}(\vec{k})C_{2,s}(\vec{k}+\vec{q})\alpha_{s}(k_{x}+q_{x}) -iC^{\dagger}_{2,s}(\vec{k})C_{1,s}(\vec{k}+\vec{q})\alpha_{s}(k_{x}+q_{x})\Big] \nonumber\\&&
W(k_{x},\vec{q})=  \frac{ 2\hat{g}s(k_{x})e^{-i tan^{-1}(\frac{q_{z}}{ 2\hat{g}s(k_{x})})}}{ \sqrt{(2\hat{g}s(k_{x}))^2+q^2_{z}}}\nonumber\\&&
\end{eqnarray}
where $\vec{q}$ is the photon momentum  and $\Omega(\vec{q})$ is the photon frequency. We introduce chiral operators $C_{R,s}(\vec{k})$, $C_{L,s}(\vec{k})$ and find the  two eigenvalues $\pm k_{y}-k_{0}$:
\begin{equation}
C_{1,s}(\vec{k})=\frac{1}{\sqrt{2}}\Big(C_{R,s}(\vec{k})+C_{L,s}(\vec{k})\Big);
 C_{2,s}(\vec{k})=\frac{i}{\sqrt{2}}\Big(C_{R,s}(\vec{k})-C_{L,s}(\vec{k})\Big)
\label{equation}
\end{equation}
In the chiral representation, the Hamiltonian $ H^{\perp}$ takes the form:
\begin{eqnarray}
&&H^{\perp}_{0}=\int\frac{d^2k}{(2\pi)^2}\sum_{s=\pm} \Big[ \hbar v( k_{y}-k_{0})C^{\dagger}_{R,s}(\vec{k})C_{R,s}(\vec{k})\alpha_{s}(k_{x})+  \hbar v(- k_{y}-k_{0})C^{\dagger}_{L,s}(\vec{k})C_{L,s}(\vec{k})\alpha_{s}(k_{x}) \Big]\nonumber\\&&
H^{\perp}_{A_{y}}=\sum_{s=\pm}\int\frac{d^2k}{(2\pi)^2}W(k_{x},\vec{q})A_{y}(\vec{q},t)\Big[C^{\dagger}_{L,s}(\vec{k})C_{L,s}(\vec{k}+\vec{q})\alpha_{s}(k_{x}+q_{x})-C^{\dagger}_{R,s}(\vec{k})C_{R,s}(\vec{k}+\vec{q})\alpha_{s}(k_{x}+q_{x}) \Big]\nonumber\\&&
\end{eqnarray}
In Eq.(10) we observe that the right chiral electrons  have the band dispersion $ \hbar v( k_{y}-k_{0})$ and the left chiral electrons  have the band dispersion  $ \hbar v(- k_{y}-k_{0})$.

In the next    stage,  we will replace the chiral operators   $ C_{R,s}(\vec{k})$ ,$ C_{L,s}(\vec{k})$ with  the particle  $a_{R,s}(\vec{k})$, $a_{L,s}(\vec{k})$ and anti-particle operators $b_{R,s}^{\dagger}(-\vec{k})$, $b_{L,s}^{\dagger}(-\vec{k})$ where $\theta[k_{y}]=1$  for $ k_{y}\geq 0 $ : 
\begin{eqnarray}
&& C_{R,s}(\vec{k})=\theta[k_{y}] \alpha_{s}(k_{x})a_{R,s}(\vec{k})+\theta[-k_{y}]\alpha_{s}(k_{x})b_{R,s}^{\dagger}(-\vec{k})\nonumber\\&&
C_{L,s}(\vec{k})=\theta[-k_{y}] \alpha_{s}(k_{x})a_{L,s}(\vec{k})+  \theta[k_{y}]\alpha_{s}(k_{x})b_{L,s}^{\dagger}(-\vec{k})\nonumber\\&&
\end{eqnarray}
This allows us  to replace the chiral  bands  in Eq.(10)  and introduce for  the boundary surface the    $ valence$ and $conduction$  bands. The representation  in Eq.(11)is essential for the use of the Wick theorem \cite{Peskins} which relies on the property that the operators for the particles   $a_{R,s}(\vec{k})$, $a_{L,s}(\vec{k})$ and anti-particle $b_{R,s}(-\vec{k})$, $b_{L,s}(-\vec{k})$    annihilate the ground state.

\vspace{0.2 in}

\textbf{III.The  vacuum solid interface Hamiltonian }

\vspace{0.2 in}

In the next section we present the vacuum-solid interface Hamiltonian. At   the interface $z=0$  we have a region   of width $d$ where both  $H^{\perp}$ and  $H^{(vacuum)}$ are valid. 
$H^{(vacuum)}$ is given by: 
\begin{eqnarray}
&& H^{(vacuum)}=\int\frac{d^{2}k}{(2\pi)^2}\int_{0}^{\infty}\frac{dk_{z}}{\pi}\sum_{\sigma=\uparrow,\downarrow}\Big[E(\vec{k},k_{z})f^{\dagger}_{\sigma}(\vec{k},k_{z})f_{\sigma}(\vec{k},k_{z})\Big],\vspace{0.02 in} E(\vec{k},k_{z})= E(\vec{k},0)+\Delta;\nonumber\\&& E(\vec{k})=\frac{\hbar^2|\vec{k}|^{2}}{2m}, \Delta=\frac{\hbar^2k^2_{z}}{2m}  \nonumber\\&&
\end{eqnarray}
$H^{(vacuum)}$ is given in terms  of  the electron operators  $f_{\sigma}(\vec{k},z)$, $ f^{\dagger}_{\sigma}(\vec{k},z)$, and  a parabolic energy dispersion  $ E(\vec{k},k_{z})= E(\vec{k},0)+\Delta $ in the   the region $ [0,L]$. The dispersion   $E(\vec{k},0)$  represents the energy  of the emitted electrons  given as a function of    the parallel conserved  momentum $\vec{k}$. The $z$  component of the momentum is integrated out.    The $z$-dependent energy is given by    $\Delta=\frac{\hbar^2k^2_{z}}{2m}$.  
The interface Hamiltonian is controlled by the vacuum solid potential $-V_{0}(z)$, which changes from zero outside the crystal to $-V_{0}$ inside the crystal. The interface Hamiltonian is approximated  by a smooth function of the form $(e^{\kappa (z-d)}-1)V_{0}$.  At $z=d$ the potential is zero and for $z=-L$ the potential is -$V_{0}$ :

\vspace{0.1 in}

\begin{eqnarray}
&&H^{(interface)}=\nonumber\\&&\int_{-L}^{d}\,dz\int\,d^2x\sum_{\sigma=\uparrow,\downarrow} f_{\sigma}^{\dagger}(\vec{x},z)\Big(e^{\kappa (z-d)}-1\Big)V_{0}\Psi(\vec{x},z)+h.c. =
\int_{0}^{\infty}\frac{dk_{z}}{\pi}\int\frac{d^{2}k}{(2\pi)^2}S(\vec{k},k_{z};\kappa)V_{0}\cdot\nonumber\\&& \Big[f_{\uparrow}^{\dagger}(\vec{k},k_{z})\Big(
\frac{\alpha_{+}(k_{x})}{\sqrt{2}}(iC_{1,+}(\vec{k})+C_{2,+}(\vec{k}))+\frac{\alpha_{-}(k_{x})}{\sqrt{2}}
(-iC_{1,-}(\vec{k})+C_{2,-}(\vec{k})\Big)\nonumber\\&&+f_{\downarrow}^{\dagger}(\vec{k},k_{z})\Big(\frac{\alpha_{+}(k_{x})}{\sqrt{2}}(C_{1,+}(\vec{k})+iC_{2,+}(\vec{k}))+\frac{\alpha_{-}(k_{x})}{\sqrt{2}}(C_{1,-}(\vec{k})-iC_{2,+}(\vec{k})\Big) +h.c.\Big]=\nonumber\\&&
\int_{0}^{\infty}\frac{dk_{z}}{\pi}\int\frac{d^{2}k}{(2\pi)^2}S(\vec{k},k_{z};\kappa)V_{0}\Big[if_{\uparrow}^{\dagger}(\vec{k},k_{z})\Big(\alpha_{+}(k_{x})C_{R,+}(\vec{k})-\alpha_{-}(k_{x})C_{L,-}(\vec{k})\Big)+\nonumber\\&&f_{\downarrow}^{\dagger}(\vec{k},k_{z})\Big(-\alpha_{+}(k_{x})C_{L,+}(\vec{k})+\alpha_{-}(k_{x})C_{R,-}(\vec{k})\Big)+h.c.\Big]\nonumber\\&&
\end{eqnarray}
where the integration with respect $z$ introduces the function $S(\vec{k},k_{z};\kappa)$ which describes the interface: 
$S(\vec{k},k_{z};\kappa)=\int_{-L}^{0}\,dz \sqrt{2\hat{g}(s(k_{x})}\Big[e^{i(k_{z}+\hat{g}(s(k_{x})z)}\cdot\Big(e^{\kappa z}-1\Big)\Big]$.
In obtaining Eq.(13), the fermion
 $f_{\sigma}(\vec{k},z)$   is replaced  by a spinor with  two  spin components and two orbitals.  For the free electrons  we will choose  the spinor  representations  $f_{\sigma=\uparrow}\rightarrow\Big[1,0,1,0\Big]^{T}$ and  $f_{\sigma=\downarrow}\rightarrow\Big[0,1,0,1\Big]^{T}$. 
  The  term   $S (\vec{k},k_{z};\kappa)$  depends on the parallel momentum $\vec{k}$  and the $k_{z}>0$ momentum which is a function of the energy  $\Delta$.

\vspace{0.2 in} 

\textbf{IV.The  detection of  photoelectrons }

\vspace{0.2 in}

The vector potential for a  photon field of  frequency $\Omega$ is given by  $\vec{A}(\vec{x},t)=\vec{e}_{r}(\vec{q})\frac{A_{r}(\vec{q})}{\sqrt{2\Omega(\vec{q})}}
e^{i\vec{q}\cdot\vec{x}}e^{-i\Omega t}+\vec{e}_{r}(\vec{q})\frac{A_{r}^{\dagger}(\vec{q})}{\sqrt{2\Omega(\vec{q})}}
e^{-i\vec{q}\cdot\vec{x}}e^{i\Omega t}$.
 The photons propagate in   the direction  $\frac{\vec{q}}{|\vec{q}|}=\Big[\sin(\theta) \cos(\phi), \sin(\theta) \sin(\phi),\cos (\theta)\Big]$  with respect to  the sample surface.  The photon field has  two orthogonal polarizations to $\frac{\vec{q}}{|\vec{q}|}$ which  are given by  $\vec{e}_{r}(\vec{q})$, $r=1,2$. The  vector potential  $\vec{A}(\vec{x},t)$ corresponds to a single photon of  frequency $\Omega$ and momentum  $\vec{q} $.

In order to compute the intensity of the emitted    photoelectrons,  we need to identify  the scattering    Hamiltonian  $ H^{ext.}_{int.}$  responsible   for the photoemission.
 $ H^{ext.}_{int.}$ is the   sum of the interface  Hamiltonian  $H^{(interface)}$ given in Eq.(13) and the  $H^{\perp}_{A_{y}}$ Hamiltonian  given in Eq.(8).
\vspace{0.1 in}
\begin{equation}
 H^{ext.}_{int.}=H^{(interface)}+ H^{\perp}_{A_{y}} 
\label{external}
\end{equation}

From the Hamiltonian  $ H^{ext.}_{int.}$ we obtain the $S$-matrix. The scattering matrix is defined as a process where  an incoming state $| i\rangle$ scatters  to an outgoing state  $| f\rangle$. We represent this process as:
\begin{equation}
_{outgoing}\langle f| i\rangle  _{incoming} =S_{f,i} = \langle f |T\Big[e^{-\frac{i}{\hbar}\int_{-\infty}^{\infty}\,dt' H^{ext.}_{int.}(t')}\Big]|i \rangle 
\label{smatrix} 
\end{equation} 
The $S$-matrix  is given as a time order product ($T$) that   acts on  the initial  state  $| i\rangle$ and final state  $| f\rangle$  \cite{Peskins}. 
In a photoemission experiment  incoming  photons produce outgoing electrons.   The $S$ matrix for an incoming photon and  an outgoing electron only  is zero. If the emitted electron is accompanied by an internal  excitation, the $S$  matrix is finite .  
 When the  $initial$ state $|  i\rangle$ is  a single photon of frequency $\Omega$ and polarization $r$,  $|  i  \rangle=A^{\dagger}_{r}(\vec{q})| 0\rangle$, and the outgoing electron is accompanied by a hole excitation in the valence band ,  at the second order in  $ H^{ext.}_{int.}$ we   find a finite $S$ matrix.
 The Hamiltonian  $ H^{ext.}_{int.}$ allows for two chiral states for the hole operators and  two spin polarizations of   the emitted electrons. 
When the  final state  $| \chi \rangle$ is   unknown,  the  scattering amplitude is represented  by  $S_{\chi,i}$, which is  approximated by  $S_{f^{(L,\sigma)},i}$, $S_{f^{(R,\sigma)},i}$ or the non-relativistic amplitude  $S_{f^{(parabolic,\sigma)},i}$. A good approximation  satisfies $|S_{\chi,i}|^2\approx\sum_{\sigma\uparrow,\downarrow}\Big(|S_{f^{(R,\sigma)},i}|^2+|S_{f^{(L,\sigma)},i}|^2+|S_{f^{(parabolic,\sigma)},i}|^2\Big)$ ( see the discussions in the next section).

The final states 
  $|  f^{(R,\sigma)} \rangle$ and$ |  f^{(L,\sigma )} \rangle$ are  given by:
\begin{eqnarray}
  &&|  f^{(R,\uparrow)} \rangle=\sum_{s=1,-1}  \Big[f^{\dagger}_{\sigma=\uparrow}(\vec{p},p_{z}) b^{\dagger}_{R,s}(-\vec{p})\alpha_{s}(p_{x})\theta[-p_{y} ]\Big]| 0\rangle   \nonumber\\&& 
|  f^{(R,\downarrow)} \rangle=\sum_{s=1,-1}  \Big[f^{\dagger}_{\sigma=\downarrow}(\vec{p},p_{z}) b^{\dagger}_{R,s}(-\vec{p})\alpha_{s}(p_{x})\theta[-p_{y} ]\Big]| 0\rangle  \nonumber\\&&
|  f^{(L,\uparrow)} \rangle=\sum_{s=1,-1}  \Big[f^{\dagger}_{\sigma=\uparrow}(\vec{p},p_{z}) b^{\dagger}_{L,s}(-\vec{p})\alpha_{s}(p_{x})\theta[p_{y}]| 0\rangle    \nonumber\\&&
|  f^{(L,\downarrow)} \rangle=\sum_{s=1,-1}  \Big[f^{\dagger}_{\sigma=\downarrow}(\vec{p},p_{z}) b^{\dagger}_{L,s}(-\vec{p})\alpha_{s}(p_{x})\theta[p_{y}]| 0\rangle  \nonumber\\&&
\end{eqnarray} 
We first consider the case in which  the final state is   $|f^{(L,\sigma)} \rangle$ 
and  obtain information on the  valence band  dispersion  $\hbar v(-k_{y}-k_{0})$.  Choosing  $|f^{(R,\sigma)} \rangle$,
we will obtain  the information for the band $\hbar v(k_{y}-k_{0})$.
To compute the matrix element  $S_{f^{(L,\sigma)},i}$  for the initial photon state $|i\rangle=A^{\dagger}_{r}(\vec{q})| 0\rangle$ and final state $|f^{(L,\sigma)} \rangle$ we use  Eq.(16).  The  second order in  $H^{ext.}_{int.}$  gives   the amplitude $S_{f^{(L,\sigma)},i}$ :
\begin{eqnarray}
&&S_{f^{(L,\sigma=\uparrow)},i}=(\frac{-i}{\hbar})^2\int_{-\infty}^{\infty}\,dt\int_{-\infty}^{\infty}\,dt'\langle 0( f^{(L,\sigma=\uparrow)})^{\dagger}|T\Big[H^{ext.}_{int.}(t)H^{ext.}_{int.}(t')\Big]A^{\dagger}_{r}(\vec{q})|0\rangle
\nonumber\\&&
S_{f^{(L,\sigma=\downarrow)},i}=(\frac{-i}{\hbar})^2\int_{-\infty}^{\infty}\,dt\int_{-\infty}^{\infty}\,dt'\langle 0( f^{(L,\sigma=\downarrow)})^{\dagger}|T\Big[H^{ext.}_{int.}(t)H^{ext.}_{int.}(t')\Big]A^{\dagger}_{r}(\vec{q})|0\rangle
\nonumber\\&&
\end{eqnarray}
The Wick theorem  reduces the $T$ -product to a product of contractions ( given by  Green's function),then it acts on  the uncontracted fields on the initial  state $|i\rangle=A^{\dagger}_{r}(\vec{q})|0\rangle$ and final state $\langle f^{(L,\sigma)}|$. The computation of the matrix element  $S_{f^{(L,\sigma=\uparrow)},i}$ is 
given in $Appendix A$.

The  result  of the Wick theorem for the final state,$\langle f^{(L,\sigma)}|$  is: one   photon is  absorbed,    one    electron is created  in the conduction band and a  hole is excited in the  valence band. 
The result for the matrix element  $S_{f^{(L,\uparrow)},i}$  in Eq.(18)  is a function of the emitted electron energy $ E(\vec{k},k_{z})$.  The information regarding  the quasi-particle excitation  is obtained from the  free electrons which have a finite  amplitude to propagate into the region $z>0$ with  the same  parallel momentum  and the  energy $ E(\vec{k},0)$.  The integration with respect to the momentum $k_{z}$  replaces   the valence electron  energy  with    $ E(\vec{k},0)$ plus an average with respect  to $\Delta(k_{z})$.:
\begin{eqnarray}
 &&S_{f^{(L,\uparrow)},i}\propto (\frac{-i}{\hbar})^2\int_{0}^{\infty}\frac{dk_{z}}{\pi}
M(\vec{k},k_{z},\vec{q},\kappa)  \theta[-k_{y}]\theta[k_{y}+q_{y}]\theta[M^2-k^2_{x}]\cdot \nonumber\\&&
\frac{-i}{\Big(E(\vec{k},0)+\Delta(k_{z})-\hbar v(k_{y}+k_{0})\theta[-k_{y}]\Big)}\cdot \delta\Big[E(\vec{k},0)+\Delta(k_{z})+\hbar v(k_{y}+q_{y}+k_{0})\theta[k_{y}+q_{y}]-\hbar \Omega(\vec{q})\Big]\nonumber\\&&
\end{eqnarray}
Here,  the matrix element $M(\vec{k},k_{z},\vec{q},\kappa)$ is given by the product of $ W(\vec{k},\vec{q})$  and $S(\vec{k},k_{z},\kappa)$ introduced in Eq.(8)  and Eq.(13):
$M(\vec{k},k_{z},\vec{q},\kappa)=W(\vec{k},\vec{q}) S(\vec{k},k_{z},\kappa) V_{0} e^{(y)}_{r}(\vec{q})$.
We perform the $ k_{z}$ integration, shift the momentum  $k_{y}\rightarrow k_{y}-q_{y}$ and introduce the  life time $\Gamma$.  In Eq.(18) we replace   $E(\vec{k}-q_{y},0)-\hbar \Omega (\vec{q} )$   with  $\epsilon$ and identify the valence electron energy  , $\epsilon=E(\vec{k}-q_{y},0)-\hbar \Omega (\vec{q} )$.
 Specifically,  we find:
\begin{eqnarray} 
 &&S_{f^{(L,\uparrow)},i}\propto (\frac{-i}{\hbar})^2
M(\vec{k},\Delta(\vec{k}),\vec{q},\kappa)  \theta[-k_{y}+q_{y}]\theta[k_{y}]\theta[M^2-k^2_{x}]\cdot \nonumber\\&&
\frac{1}{\sqrt{\Big(\epsilon+\hbar v(k_{y}+k_{0})\theta[k_{y}]+i\Gamma\Big)}}\cdot  \frac{1}{\Big(\hbar\Omega(\vec{q})- \hbar v(k_{y}+q_{y}+k_{0})+i\Gamma\Big)}\nonumber\\&&
\end{eqnarray}

This expression demonstrate    that  the amplitude $S_{f^{(L,\uparrow)},i}$ has information about   the valence band  dispersion   
$\epsilon=\hbar v(-k_{y}-k_{0})$ . Not shown is the amplitude   $S_{f^{(R,\uparrow)},i}$ with   the dispersion $\epsilon=\hbar v(k_{y}-k_{0})$.  In Eq.(19) we have introduced  the life time $ \Gamma$ which is a result of electron-electron interactions (see section $ VI$) or impurities scattering. The elastic impurities scattering  is less significan  since  the Hamiltonian  $H^{\perp}_{0}$ ( Eq. 10) is time reversal invariant $\hat{T}$ in two space dimensions,    obeing the condition $\hat{T}^2=-1$.   Our model with a disorder potential    belongs to the symplectic class   which has no localized phase \cite{Hikami}.

\noindent
  In section $VI$ we  consider   the Coulomb interaction in the absence of impurities scattering.  A number of authors have shown that at finite temperatures    the impurity scattering mean free path $ l_{imp.}$  can be larger   than the electron-electron scattering  mean free path $l_{e-e}$ , $l_{e-e}< l_{imp.}$ \cite{lucas}.  As a result as observed in graphene, that  the velocity $v$  and the life time $\Gamma$ renormalizes,     becoming temperature dependent.  It  was shown  \cite{son} that   at $T=0$ the dispersion  in  graphene is anomalous therefore we expect similar feature for    the Weyl semimetal.

\vspace{0.2 in}

\textbf{V.Physical information obtained   from the scattering matrix  }

\vspace{0.2 in}

 In the literature,   the computation of the   optical Joint density of states involving the valence band and conduction band   \cite{Yazdani}  provides information about  energy dispersion.  Our computation given in Eq.(19) reveals similar features. However, we have only one conduction  and one valence band 
giving rise to a simplified picture, as a function of the photon frequency,  shifted  energy $\epsilon=E(\vec{k}-q_{y},0)-\hbar \Omega (\vec{q} )$ and momentum $ \vec{k}$.

In a photoemission experiment the experimentalist measures the energy of the emitted electrons  $E_{em.}(\theta,\phi)$   as a function of the orientation of the crystal surface.
From this energy and orientation (using the model of free electrons) we determine $E(\vec{k},0)=E_{em.}(\theta,\phi)\sin^2(\theta)$   and $ \vec{k}(\theta,\phi)$. These results  are used to determine the valence band  energy given by  $\epsilon=  E(\vec{k},0) -  \hbar \Omega(\vec{q})=E_{em.}(\theta,\phi)\sin^2(\theta)-\hbar \Omega(\vec{q})$ used  in Eq(19).  We  generate  a plot of the experimental results  and determine if the points $(\epsilon,\vec{k})$, fit the dispersion $\epsilon=\hbar v(-k_{y}-k_{0})$  ,  $\epsilon=\hbar v(k_{y}-k_{0})$  or  $ \epsilon = \frac{\hbar^2}{2m}|\vec{k}|^2-V_{0}$ (parabolic  dispersion). In all these cases we need  to consider  the energy or temperature dependence of the life time $\Gamma$.   The dispersion $\epsilon=\hbar v(-k_{y}-k_{0})$  corresponds to the scattering  amplitude $S_{f^{(L,\uparrow)},i}$, the dispersion  $\epsilon=\hbar v(k_{y}-k_{0})$  corresponds to  $S_{f^{(R,\uparrow)},i}$ and the case $ \epsilon = \frac{\hbar^2}{2m}|\vec{k}|^2-V_{0}$  corresponds to a  non-Dirac  scattering amplitude $S_{f^{(parabolic,\sigma)},i}$. 
 The unknown scattering amplitude    $ S_{\chi,i}(\epsilon,\vec{k}) = \langle \chi |T\Big[e^{-\frac{i}{\hbar}\int_{-\infty}^{\infty}\,dt' H^{ext.}_{int.}(t')}\Big]|i \rangle$   for  the final  state $\langle\chi|$ obeys the relation  $| S_{\chi,i}(\epsilon,\vec{k})|^2\approx \sum_{\sigma=\uparrow,\downarrow}\Big(|S_{f^{(L,\sigma)},i}|^2+|S_{f^{(R,\sigma)},i}|^2+|S_{f^{(parabolic,\sigma)},i}|^2\Big)$. 

Next, we analyze the situation for the chiral final states given in $Eq.(16)$.
We   compute  $S_{f^{(L,\sigma)},i}$ and  find the following  scattering intensity : 

\vspace{0.003 in}

\begin{eqnarray}
&&|S_{f^{(L,\uparrow)},i}+S_{f^{(L,\downarrow)},i}|^2 \approx 
\nonumber\\&& \theta[-k_{y}+q_{y}]\theta[k_{y}] \frac{(2\hat{g}s(k_{x}))^2}{q^2_{z} +(2\hat{g}s(k_{x}))^2}| M(\vec{k},\Delta(\vec{k}),\vec{q},\kappa) |^{2}\frac{1}{ \sqrt{\Big((\epsilon+\hbar v(k_{y}+k_{0})\theta[k_{y}])^2+\Gamma^2\Big)}}\cdot\nonumber\\&&  \frac{1}{\Big((\hbar\Omega(\vec{q})- \hbar v(k_{y}+q_{y}+k_{0}))^2+\Gamma^2\Big)}\Big[\delta_{\sigma=\uparrow}\theta[M^2-k^2_{x}] +\delta_{\sigma=\downarrow}\theta[k^2_{x}-M^2]\Big] 
\nonumber\\&&
\end{eqnarray}
 
We observe that the scattering amplitude  $|S_{f^{(L,\sigma)},i}|^2$ has  two  singularities . The first  is given by the  shifted   valence band dispersion  $\epsilon=\hbar v(-k_{y}-k_{0})$  where  $\epsilon=E(\vec{k}-q_{y},0)-\hbar \Omega (\vec{q} )$, the second gives information on the conduction band  and photon frequency $\hbar \Omega (\vec{q})=\hbar v(k_{y}+k_{0}-q_{y})$. 
These results confirm  that both the dispersion and  the photon absorption   from  the valence to  the conduction band  can be obtained  theoretically   in  agreement with the  photoemission experiments.

Using the results given in Eq.(20),  we plot the scattering  intensity matrix. We will use the units of ev. for the momentum $k_{y}$, $k_{0}$ and frequency $\Omega$.  The intensity   $|S_{f^{L,\sigma},i}|^2$  and  the photon absorption are plotted  in  arbitrary units.

The scattering   intensity  $|S_{f^{(L,\uparrow)},i}|^2(\epsilon)$ for a   fixed momentum $k_{x}=0$ and photon frequency   $\Omega =50ev$,    reveals that    the maximum   intensity varies   with  the valence band energy $\epsilon$ and momentum $k_{y}$.

 The  scattering   intensity   $|S_{f^{(L,\sigma)},i}|^2(k_{x},k_{y})$ is  a function of the momentum   $k_{x}$ , $k_{y}$ and    shows evidence    of the nodal points at $k_{x}=\pm 0.5$.  

Figure $1$ shows the   scattering  intensity $|S_{f^{(L,\sigma=\uparrow)},i}|^2(k_{y},\epsilon)$  for photoelectrons  with spin $ \sigma=\uparrow$ and left chiralty for the valence band   $\epsilon=\hbar v(-k_{y}-k_{0})\theta[k_{y}]$ . The intensity is a function of the momentum $k_{y}$ and energy $\epsilon(\vec{ k} )$ for  the photon frequency measured  in ev.,  $\Omega =100 ev$ . This figure reveals the dispersion   $\epsilon=\hbar v(-k_{y}-k_{0})$  for $k_{y}>0$  of the valence band shifted by the laser frequency presented in the $(\epsilon,k_{y})$ plane.
\begin{figure}
\begin{center}
\includegraphics[width=4.5 in]{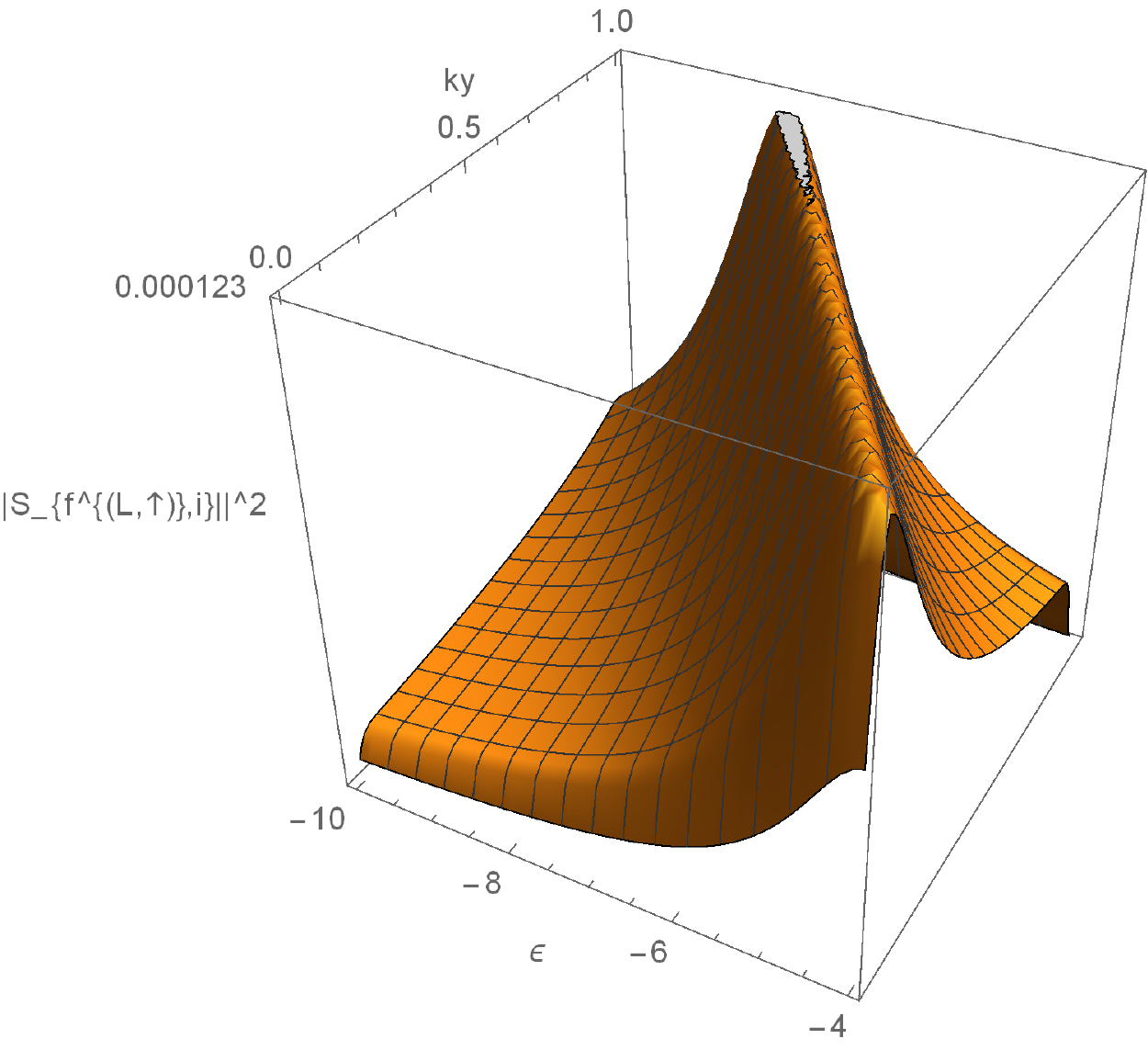}
\end{center}
\caption{The scattering  intensity  $|S_{f^{(L,\uparrow)},i}|^2(k_{y},\epsilon)$ for dispersion   $\epsilon=\hbar v(-k_{y}-k_{0})\theta[k_{y}]$  ($\epsilon=E(k_{y}-q_{y},k_{x}=0)-\hbar \Omega (\vec{q} )$).  A  life time $\Gamma$ for the valence band $ \epsilon $ was used in the plot. The confinement potential is $ V_{0}=4ev$ and the  photon  frequency is $\Omega =100 ev$. The coordinates $(x,y)$ corresponds to $(k_{y}, \epsilon)$, the $z$ axis is represented by $|S_{f^{(L,\uparrow)},i}|^2$ computed for  a small value of  the  life time  $\Gamma$.}
\end{figure}

Figure $2$ shows the  scattering intensity $|S_{f^{(L,\sigma=\uparrow)},i}|^2(k_{y},\Omega)$ as a function of the photon frequency  $\Omega$ and the momentum $k_{y}$. This figure reveals  the conduction  band  dispersion due to the  photon  absorption   $ \hbar\Omega(\vec{q})\approx \hbar v(k_{y}+k_{0})$.
\begin{figure}
\begin{center}
\includegraphics[width=4.5 in]{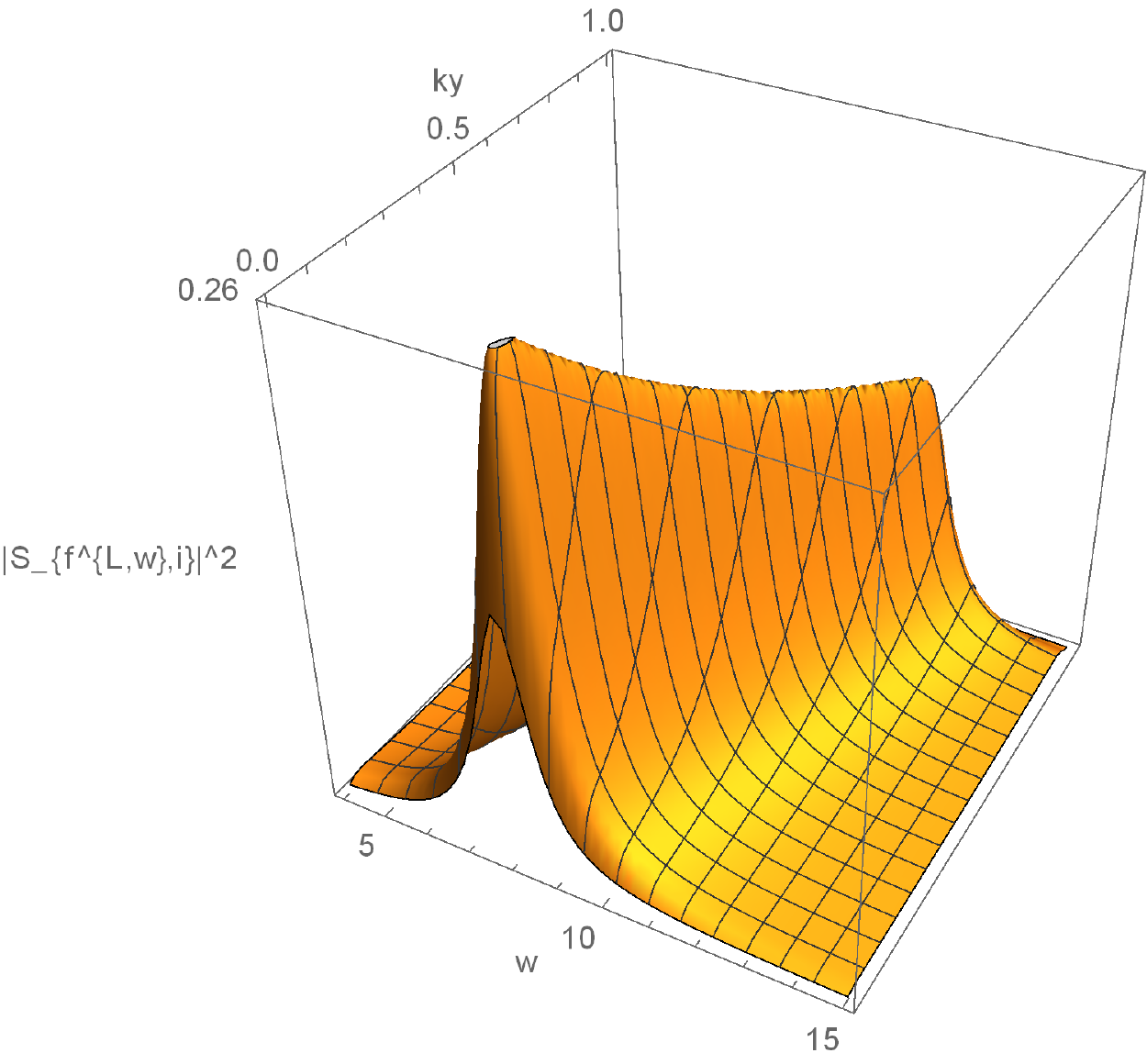}
\end{center}
\caption{The  scattering     intensity $|S_{f^{(L,\sigma\uparrow)},i}|^2(k_{y},\Omega)$ as a function of the photon  frequency  $\Omega$ and the momentum $k_{y}$. (In figure $2$ we have used the symbol $W$ for the frequency  $\Omega$.) The confinement potential is $ V_{0}=4ev$. The coordinates $(x,y)$ corresponds to $(k_{y}, \Omega)$, the $z$ axis is represented by  $|S_{f^{(L,\sigma\uparrow)},i}|^2(k_{y},\Omega)$  computed for  a small value of  the  life time  $\Gamma$.} 
\end{figure}

Figures $1$  and $2$  with the dispersion   equation $ \epsilon(\vec{k})=\hbar v(-k_{y}-k_{0})$ and  $\hbar\Omega(\vec{q})\approx  \hbar v(k_{y}+k_{0})$ are in agreement  with the experimental observations \cite{BQ}.

Figure $3$ shows the $^{''}$ Fermi arc $^{''}$ which connects the points   $M$ to  $-M$ given by the largest contour . Since the momentum in the $z$ direction is not conserved, we can plot the projected contours   on the surface $z=0$.   We consider the   contour plot   for different intensities  $|S_{f^{(L,\sigma=\uparrow)},i}|^2(k_{x},k_{y})$  as a function of the momentum $\vec{k}$ for  a fixed  valence band dispersion energy $\epsilon$ and spin polarization $ \sigma=\uparrow$. We use the $contour$ $plot $ $program$ where   the $^{''}$ Fermi arc $^{''}$  is given by the  contour plot    $|S_{f^{(L,\sigma=\uparrow)},i}|^2(k_{x},k_{y})=0$ representing the  scattering  intensity at a fixed energy. Such  a contour connects  the points  $M$ to  $-M$.  In  practice, we can plot only  contours which are close to zero. The lowest intensity  contour is   $|S_{f^{(L,\sigma=\uparrow)},i}|^2(k_{x},k_{y})=$constant$ \cdot 10^{-7}$ ,which represents the largest contour.The rest of the contours correspond  to  increasing   intensities.
$|S_{f^{(L,\sigma =\uparrow)},i}|^2(k_{x},k_{y})=$constant$ \cdot   2\cdot 10^{-7}$, $|S_{(f^{L,\sigma =\uparrow)},i}|^2(k_{x},k_{y})=$constant$ \cdot 3 \cdot 10^{-7}$  and $|S_{f^{(L,\sigma=\uparrow)},i}|^2(k_{x},k_{y})=$constant$\cdot 4 \cdot 10^{-7}$,  $|S_{(f^{(L,\sigma=\uparrow)},i}|^2(k_{x},k_{y})=$constant$\cdot 5\cdot 10^{-7}$, $|S_{f^{(L,\sigma=\uparrow)},i}|^2(k_{x},k_{y})=$constant$\cdot 6 \cdot 10^{-7}$, $|S_{(f^{(L,\sigma=\uparrow},i}|^2(k_{x},k_{y})=$constant$\cdot 7 \cdot 10^{-7}$, $|S_{f^{L,\sigma=\uparrow)},i}|^2(k_{x},k_{y})=$constant$\cdot 8 \cdot 10^{-7}$. The plots  are for a fixed energy   $\epsilon(\vec{k}) =100ev$.  The $constant$  in the contour plot  originates from the proportionality factor  $| M(\vec{k},\Delta(\vec{k}),\vec{q},\kappa) |^{2}$in Eq.(20).
\clearpage
\begin{figure}
\begin{center}
\includegraphics[width=4.5 in]{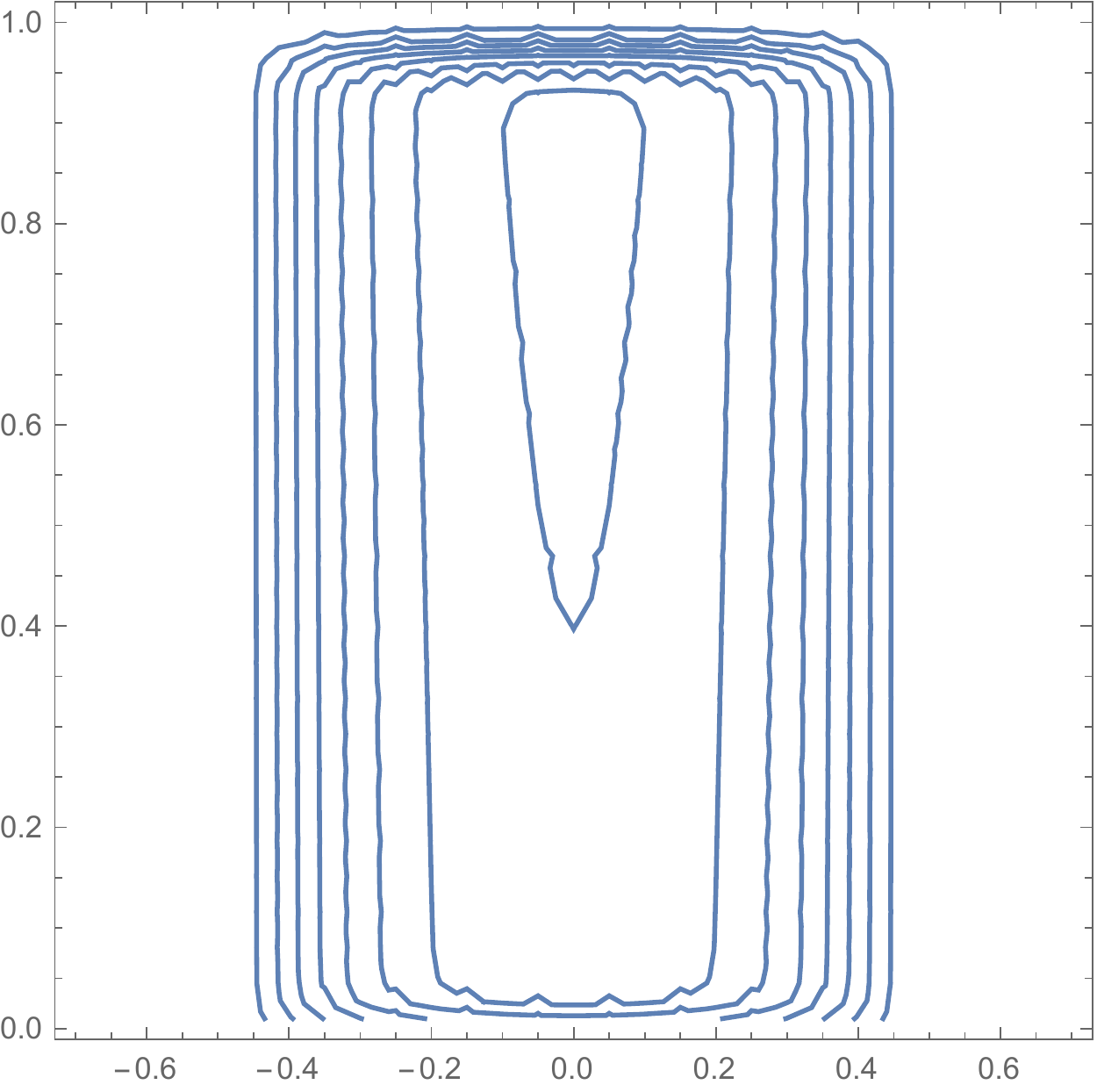}
\end{center}
\caption{The  intensity contour  as a function  of $ k_{x}$,$k_{y} $ (horizontal   and  vertical  vertical axes).  We plot  the set  solutions  of the equation   $|S_{f^{(L,\sigma=\uparrow)},i}|^2(k_{x},k_{y})=$constant$ \cdot 10^{-7}$ ( the larger   contour ) will corespond to the  $^{''}$ Fermi arc $^{''}$.  $|S_{f^{(L,\sigma\uparrow)},i}|^2(k_{x},k_{y})=$constant$\cdot 8\cdot 10^{-7}$ (the smallest  contour ) is depicted    in ascending order for a fixed energy  $\epsilon=100ev.$ }
\end{figure}
\clearpage
The intensity contour plot for $|S_{f^{(L,\sigma=\downarrow)},i}|^2(k_{x},k_{y})$ reveals the arcs     which  connect  $k_{x}=0.5$ to   $k_{x}=-0.5 $ (shown in the figure $3$) and arcs  which start at $ k_{x}=\pm0.5$  and go to large momentum  $k_{x}=\pm\infty$  which have opposite curvature (not shown).   

 By  repeating  the calculation for the final  state  $|f^{(R,\sigma)} \rangle$,
 we can obtain the scattering amplitude  for  the  valence    band  energy $\hbar v(k_{y}-k_{0})$.

 The natural question which  one can ask is what is the effect of the bulk electrons and how  can they  be detected? The bulk electrons  have a non-relativistic dispersion contrary to the Weyl fermions. 
 To detect the bulk   electrons  we need to consider the scattering matrix for  the    non- relativistic Hamiltonian  $\vec{A}\cdot \vec{p}$ following  the standard theory for photoemission \cite{Mahan}. This leads us to suggest that a realistic model for the $WSM$ needs to contain both,  linear  dispersion for the Dirac like bands  and parabolic  dispersion for the bulk bands.
The two contributions are additive and can be handled independently. According to  the discussion  in section $IV$ the non-relativistic amplitude  $S_{f^{(parabolic,\sigma)},i}$ needs to be computed   and included in the  scattering intensity   must be given as  $\approx\sum_{\sigma\uparrow,\downarrow}\Big(|S_{f^{(R,\sigma)},i}|^2+|S_{f^{(L,\sigma)},i}|^2+|S_{f^{(parabolic,\sigma)},i}|^2\Big)$. The amplitudes   will have complimentary  contributions,  the region $\epsilon=\hbar v(\pm k_{y}-k_{0})$ and $\epsilon'= \frac{\hbar^2}{2m}|\vec{k'}|^2-V_{0}$ correspond to   different energies.

\vspace{0.2 in}

\textbf{VI. The electron-electron interaction effects on the chiral  modes which propagate on the boundary}

\vspace{0.2 in}

In order to take into account the relaxations and screening on the boundary state electrons and holes we have to consider the Coulomb interactions.The first step is to project the electron operator $\hat{\Psi}_{\sigma,\tau_{0}}(\vec{k},z)$, where $ \sigma=\uparrow,\downarrow$ and  $ \tau_{0}=1,2$ into the chiral  modes  given in Eqs.(6,9).
\begin{eqnarray}
&&\Psi_{\sigma=\uparrow,\tau_{0}=1}(\vec{x},z)=\nonumber\\&&\int\frac{d^{2}k}{(2\pi)^2} \sqrt{2\hat{g}s(k_{x})}e^{\hat{g}s(k_{x})z}e^{i\vec{k}\cdot \vec{x}}\Big[\frac{i}{2}\Big(C_{R,-}(\vec{k}) +C_{L,-}(\vec{k})\Big)\alpha_{-}(k_{x})-\frac{i}{2}\Big(C_{R,+}(\vec{k}) +C_{L,+}(\vec{k})\Big)\alpha_{+}(k_{x})\Big]\nonumber\\&&
\Psi_{\sigma=\downarrow,\tau_{0}=1}(\vec{x},z)=\nonumber\\&&\int\frac{d^{2}k}{(2\pi)^2} \sqrt{2\hat{g}s(k_{x})}e^{\hat{g}s(k_{x})z}e^{i\vec{k}\cdot \vec{x}}\Big[\frac{-1}{2}\Big(C_{R,-}(\vec{k}) -C_{L,-}(\vec{k})\Big)\alpha_{-}(k_{x})-\frac{1}{2}\Big(C_{R,+}(\vec{k}) -C_{L,+}(\vec{k})\Big)\alpha_{+}(k_{x})\Big]\nonumber\\&& \Psi_{\sigma=\uparrow,\tau_{0}=2}(\vec{x},z)=\nonumber\\&&\int\frac{d^{2}k}{(2\pi)^2} \sqrt{2\hat{g}s(k_{x})}e^{\hat{g}s(k_{x})z}e^{i\vec{k}\cdot \vec{x}}\Big[\frac{i}{2}\Big(C_{R,+}(\vec{k}) -C_{L,+}(\vec{k})\Big)\alpha_{+}(k_{x})+\frac{i}{2}\Big(C_{R,+}(\vec{k}) -C_{L,-}(\vec{k})\Big)\alpha_{-}(k_{x})\Big]\nonumber\\&&\Psi_{\sigma=\downarrow,\tau_{0}=2}(\vec{x},z)=\nonumber\\&&\int\frac{d^{2}k}{(2\pi)^2} \sqrt{2\hat{g}s(k_{x})}e^{\hat{g}s(k_{x})z}e^{i\vec{k}\cdot \vec{x}}\Big[\frac{1}{2}\Big(C_{R,+}(\vec{k}) +C_{L,+}(\vec{k})\Big)\alpha_{+}(k_{x})+\frac{1}{2}\Big(C_{R,-}(\vec{k}) +C_{L,-}(\vec{k})\Big)\alpha_{-}(k_{x})\Big]\nonumber\\&&
\end{eqnarray}
 The interaction Hammiltonian is obtained with  the  
 help of the Poisson equation $\nabla^{2}a_{0}(\vec{x},z)=-J_{0}(\vec{x},z)$ for the scalar potential $a_{0}$ and electronic density  $J_{0}(\vec{x},z)$. Using the momentum representation given in Eq.(7) with the projected spinors given in Eq.(21) we find:
\begin{eqnarray}
&& J_{0}(\vec{q},z)= \nonumber\\&&\sqrt{2\hat{g}s(k_{x})}e^{\hat{g}s(k_{x})z}\sum_{s=\pm}\int\,\frac{d^2k}{(2\pi)^2}\Big[C^{\dagger}_{R,s}(\vec{k})C_{R,s}(\vec{k}-\vec{q}) +C^{\dagger}_{L,s}(\vec{k})C_{L,s}(\vec{k}-\vec{q}) \Big]\alpha_{s}(k_{x})\alpha_{s}(k_{x}-q_{x})\nonumber\\&&
\end{eqnarray}
The Coulomb ineraction emerges from the integration of the scalar field $a_{0}(\vec{x},z)$, given in $Appendix B$.

\noindent
As a result of the Coulomb interaction $H^{e-e}$ the Hamiltonian $H^{\perp}$ is replaced by $ \tilde{ H}^{\perp}$:
\begin{equation} 
\tilde{ H}^{\perp}=H^{\perp}+ H^{e-e}
\label{hamiltonian}
\end{equation}

Computing  the scaling dimension, we observe  that  the Hamiltonian $ H^{e-e}$ is   a marginal operator.This we can see in the following way by introducing the ultraviolet cutoff $\Lambda$. We scale    $\Lambda\rightarrow \frac{\Lambda}{s}$, with $s>1$. The re-scaling of the cutoff is achieved by   the transformation of the  momentum, $k'=sk$,  $p'=sp$, $q'=sq$ and frequency $\omega'=s\omega$. The scaling of the momentum induces the  scaling for  the operators
$C'_{R,s}(\vec{k'})=s^{-\beta}C_{R,s}(\vec{k})$, $C'_{L,s}(\vec{k'})=s^{-\beta}C_{L,s}(\vec{k})$ \cite{Shankar}.
In addition, we need to check the scaling of   $ \alpha_{+}(k_{x}) $  and  $ \alpha_{-}(k_{x}) $.  In the low momentum limit only  $ \alpha_{-}(k_{x}) $ remains invariant.   Using scaling $ k'_{x}=sk_{x}$  with $s>1$ shows that if  $k^2_{x}<M^2 $  the scaling relation is satisfied  automatically  $k^2_{x}s^{-2}<M^2 $. This will not be the case for $ \alpha_{+}(k_{x}) $.  The parameter  $s(k_{x})=|[(\frac{k_{x}}{M})^2-1]|$  is replaced under scaling  by  $s(k'_{x})=|[(\frac{k_{x}}{M s})^2-1]|\approx 1$ which can be taken in the long wave limit to be constant . 
This analysis simplifies the Hamiltonian in the long wave approximation replacing  $\tilde{H}^{\perp}$ with the long wave form $H_{eff.}$.
\begin{eqnarray}
&&H_{eff.}=\int\frac{d^2k}{(2\pi)^2} \Big[ \hbar v( k_{y}-k_{0})C^{\dagger}_{R}(\vec{k})C_{R}(\vec{k})+  \hbar v(- k_{y}-k_{0})C^{\dagger}_{L}(\vec{k})C_{L} \Big]+ \int\,\frac{d^2k}{(2\pi)^2}\int\,\frac{d^2p}{(2\pi)^2}\int\,\frac{d^2q}{(2\pi)^2}\nonumber\\&&
 \cdot\Big[C^{\dagger}_{R}(\vec{k})C_{R}(\vec{k}-\vec{q}) +C^{\dagger}_{L}(\vec{k})C_{L}(\vec{k}-\vec{q}) \Big]  V^{c}(|\vec{q}|)  \Big[C^{\dagger}_{R}(\vec{p})C_{R}(\vec{p}+\vec{q}) +C^{\dagger}_{L}(\vec{p})C_{L}(\vec{p}+\vec{q})\Big]\nonumber\\&&
+\int\frac{d^2k}{(2\pi)^2}W(k_{x},\vec{q})A_{y}(\vec{q},t)\Big[C^{\dagger}_{L}(\vec{k})C_{L}(\vec{k}+\vec{q})-C^{\dagger}_{R}(\vec{k})C_{R}(\vec{k}+\vec{q}) \Big]\nonumber\\&&
W(k_{x},\vec{q})=  \frac{ 2\hat{g}e^{-i tan^{-1}(\frac{q_{z}}{ 2\hat{g}})}}{ \sqrt{(2\hat{g})^2+q^2_{z}}}\nonumber\\&&
\end{eqnarray}
The boundary surface  for  a two dimensional  space gives  rise to a $2+1$ scaling problem. The Hamiltonian $H^{\perp}_{0}$ is scale invariant  and fixes $\beta$ to  $\beta=-2$. As a result we find that the potential $V^{c}(|\vec{q}|)= \frac{e^2}{2\epsilon |\vec{q}|}$   is marginal, $s^0$ \cite{Shankar}.Our surface boundary model  with interactions is in the same universality class as  interacting graphene \cite{lucas}. We  find that the Coulomb potential is marginal irrelevant  (see  Appendix B ) $ V^{c}(|\vec{q}|)=\frac{\alpha_{0}}{2\epsilon  |\vec{q}|}$.Due to renormalization effects  
the  potential is replaced by:
 $V^{c}(|\vec{q}|)=\frac{\alpha_{eff}}{2  |\vec{q}|}$, with  $\alpha_{eff}=\alpha_{0}\Big(1+\frac{\alpha_{0}}{4}Ln\frac{\Lambda}{T}\Big)^{-1}$. We observe  that the potential is vanishing logarithmically at low temperature,
 the velocity $v$ and  the life time $\Gamma$ become temperature  dependent:
\begin{equation}
v(T)=v\Big(1+\frac{\alpha_{0}}{4}Ln\frac{\Lambda}{T}\Big),\vspace{0.1 in} \Gamma(T)\approx \alpha^2_{0}K_{B}T
\label{ parameter}
\end{equation}

\vspace{0.2 in}

\textbf{VII.  Conclusions}

\vspace{0.2 in}

We have    introduced a model  for the $WSM$  with two nodal points and   a surface boundary at $z=0$, giving rise to two chiral bands. The model uses a Hamiltonian   which considers a  non-linear  function  connecting   the nodes. Using  the minimal coupling principle, we couple the electrons  to photons without relying on the Foldy-Wouthuysen transformation.
Thus  traditional non-relativistic coupling $ \vec{A}\cdot \vec{p}$ is avoided allowing for an exact modeling of electrons and  photons. The Coulomb interaction  is  expressed in terms of  the    chiral  surface modes. We find that the projected interaction   gives rise to a model  which is in the same universality class as graphene in $2+1$ dimensions, resulting in a normalization of the  velocity and the life time.

\noindent
The main goal  of this work is to introduce  the $S$ matrix for   investigating   photoemission. We find that our theory  explains  most of the experimental observations.  The limitation of one band, two nodes  versus many bands and nodes in a real $WSM$ does  not to change the photoemission  profile.   Our  model reveals the special  properties  of the $WSM$ observed experimentally.  
 We  theoretically  compute  the valence and conduction  band dispersion for the chiral surface boundary  and demonstrate   the  emergence  of the $^{''}$Fermi arcs$^{''}$.
   When  the number of nodes is  larger  than two, such as  in the $TaAs$, the methodology used in this paper can be applied  by    replacing  the quadratic function   $(k^2_{x}-M^2)$ with  a polynomial function  $f(k_{x})$ which has   $2N$ zeros. This will give rise to multiple valence band dispersions  for different crystal surfaces. 

\vspace{0.2in}

\textbf{Appendix A}

\vspace{0.1in}

The Wick theorem  reduces the $T$ -product to a product of contractions ( given by  Green's function),then it acts on  the uncontracted fields on the initial  state $|i\rangle=A^{\dagger}_{r}(\vec{q})|0\rangle$ and final state $\langle f^{(L,\sigma=\uparrow)}|$, $\langle f^{(L,\sigma=\downarrow)}|$.
  For  $\langle f^{(L,\sigma=\uparrow)}|$,  Wick theorem generates the following result:
\begin{eqnarray}
&& S_{f^{(L,\uparrow)},i}\propto (\frac{-i}{\hbar})^2\int_{0}^{\infty}\frac{dk_{z}}{\pi}
M(\vec{k},k_{z},\vec{q})  \theta[-k_{y}]\theta[k_{y}+q_{y}] \int_{-\infty}^{\infty}\,dt\int_{-\infty}^{\infty}\,dt' (-i)\cdot\nonumber\\&& \langle f^{(L,\sigma=\uparrow)}|\Big(\theta[-M^2+k^2_{x}] f^{\dagger}_{\uparrow}(\vec{k},k_{z};t)a_{L,s=-}(\vec{k};t)  a^{\dagger}_{L,s=-}(\vec{k};t')  b_{L,s=-}(-k_{y}-q_{y};t')e^{-i\hbar\Omega(\vec{q})t'}A_{r}(\vec{q})\Big)A^{\dagger}_{r}(\vec{q})|0\rangle 
 \nonumber\\&&
\propto (\frac{-i}{\hbar})^2\int_{0}^{\infty}\frac{dk_{z}}{\pi}
M(\vec{k},k_{z},\vec{q})  \theta[-k_{y}]\theta[k_{y}+q_{y}] \theta[-M^2+k^2_{x}]\int_{-\infty}^{\infty}\,dt\int_{-\infty}^{\infty}\,dt'\cdot\nonumber\\&&\Big(e^{-iE(\vec{k},k_{z})t}\langle T(a_{L,s=-}(\vec{k};t)  a^{\dagger}_{L,s=-}(\vec{k};t') )\rangle  e^{i\hbar v(k_{y}+k_{0}+q_{y})t'}e^{-i\hbar\Omega(\vec{q})t'}\Big)\nonumber\\&&
\end{eqnarray}
 From Wick theorem, we obtain the  following   the scattering amplitude  for $\langle f^{(L,\sigma=\downarrow)}|$:
\begin{eqnarray}
&& S_{f^{(L,\downarrow)},i}\propto (\frac{-i}{\hbar})^2\int_{0}^{\infty}\frac{dk_{z}}{\pi}
M(\vec{k},k_{z},\vec{q})  \theta[-k_{y}]\theta[k_{y}+q_{y}] \int_{-\infty}^{\infty}\,dt\int_{-\infty}^{\infty}\,dt\nonumber\\&& \langle f^{(L,\sigma=\downarrow)}|\Big(\theta[M^2-k^2_{x}]   f^{\dagger}_{\downarrow}(\vec{k},k_{z};t)a_{L,s=+}(\vec{k};t) a^{\dagger}_{L,s=+}(\vec{k};t')b_{L,s=+}(-k_{y}-q_{y};t')e^{-i\hbar\Omega(\vec{q})t'}A_{r}(\vec{q})\Big)A^{\dagger}_{r}(\vec{q})|0\rangle \nonumber\\&&
\propto (\frac{-i}{\hbar})^2\int_{0}^{\infty}\frac{dk_{z}}{\pi}
M(\vec{k},k_{z},\vec{q},\kappa)  \theta[-k_{y}]\theta[k_{y}+q_{y}]\theta[M^2-k^2_{x}] \int_{-\infty}^{\infty}\,dt\int_{-\infty}^{\infty}\,dt'\cdot\nonumber\\&&
\Big(e^{iE(\vec{k},k_{z})t}\langle T(a_{L,s=+}(\vec{k};t)  a^{\dagger}_{L,s=+}(\vec{k};t') )\rangle  e^{i\hbar v(k_{y}+k_{0}+q_{y})t'}e^{-i\hbar\Omega(\vec{q})t'}\Big)\nonumber\\&&
\end{eqnarray}
Here  the matrix element $M(\vec{k},k_{z},\vec{q},\kappa)$ is given by the product of $ W(\vec{k},\vec{q})$   and $S(\vec{k},k_{z},\kappa)$ introduced in Eq.(8)  and Eq.(13):
$M(\vec{k},k_{z},\vec{q},\kappa)=W(\vec{k},\vec{q}) S(\vec{k},k_{z},\kappa) V_{0} e^{(y)}_{r}(\vec{q})$.

\vspace{0.2 in}

\textbf{Appendix B}

\vspace{0.2 in}

 We solve the Poisson equation. We integrate   the scalar field $a_{0}(\vec{x},z)$ and find:
\begin{eqnarray}
&&H^{int.C}=\nonumber\\&&\int\,\frac{d^2q}{(2\pi)^2}\int\,\frac{dq_{z}}{2\pi}\Big[\frac{1}{2\epsilon e^2}a_{0}(\vec{q},q_{z})\Big(q^2_{x}+q^2_{y}+q^2_{z}\Big)a_{0}(-\vec{q},-q_{z})+a_{0}(\vec{q},q_{z})\Big(\int_{-L}^{0}\,dz\sqrt{2\hat{g}s(k_{x})}e^{(\hat{g}s(k_{x})+iq_{z})z}\Big)\nonumber\\&&\cdot\sum_{s=\pm}\int\,\frac{d^2k}{(2\pi)^2}\Big(C^{\dagger}_{R,s}(\vec{k})C_{R,s}(\vec{k}-\vec{q}) +C^{\dagger}_{L,s}(\vec{k})C_{L,s}(\vec{k}-\vec{q}) \Big)\alpha_{s}(k_{x})\alpha_{s}(k_{x}-q_{x})\Big]\nonumber\\&&
\end{eqnarray}
The scalar field correlation is given by the longitudinal  Coulomb propagator $ D_{0}(\vec{q}, q_{z})=\frac{e^2}{2\epsilon(q^2_{x}+q^2_{y}+q^2_{z})}$.
The integration with respect to the  $k_{z}$ momentum in the presence of the second term in Eq.28  and  the long wave limit $ s(k_{x})\approx 1 $ gives  the effective surface potential  $ V^{c}(|\vec{q}|)$:
\begin{eqnarray}
&&V^{c}(|\vec{q}|)=\int\,\frac{dq_{z}}{2\pi}\frac{2 e^2}{\epsilon(q^2_{x}+q^2_{y}+q^2_{z})}\cdot\frac{1}{1+\frac{q^2_{z}}{2\hat{g}}}\approx \frac{e^2}{2\epsilon |\vec{q}|}+  \frac{e^2}{2\pi\epsilon}\cdot \frac{\Lambda}{2\hat{g}}\Big[1+\frac{|\vec{q}|}{\Lambda}Ln\Big(\frac{|\vec{q}|}{\Lambda}\Big)...\Big]\nonumber\\&&
\end{eqnarray}
where  $V^{c}(|\vec{q}|)$ is marginal and the higher order terms give  irrelevant contributions  \cite{Shankar}. The correction term $\frac{\Lambda}{2\hat{g}}$ will give rise to a  shift of  the  crystal -vacuum potential $V_{0}$ introduced in Eq.(2).
Thie  strength  of the interactions is characterized by   $\alpha_{0}=\frac{1}{137}\frac{c}{\epsilon_{r}v}$ where  $\epsilon_{0}\epsilon_{r}$ is the dielectric constant.   The   leading order   potential is  given by  $ V^{c}(|\vec{q}|)=\frac{\alpha_{0}}{2  |\vec{q}|}$.

\end{document}